\documentclass[10pt, aps, prl, superscriptaddress, nofootinbib,
               amsmath, 
               twocolumn, 
               showpacs, 
               floatfix, 
               ]{revtex4-2}

\providecommand{\exclude}[1]{}
\usepackage{etoolbox}
\newtoggle{mypaper}
\toggletrue{mypaper}

\iftoggle{mypaper}{
  \usepackage[aps]{my_paper}   
  \iftoggle{arXiv}{
  }{
    
  }
}{
  \usepackage{graphicx}
  \usepackage{booktabs}
  \graphicspath{{./figures/aps/}{./figures/}}

  \usepackage{siunitx}
  \sisetup{
    round-mode = none,   
    mode = text,         
    detect-all = true,            
    detect-mode = false,          
    list-units = single,  
    unit-mode = text,
    number-mode = math,
  }

  \usepackage[nameinlink]{cleveref}
  \providecommand{\mycaptionhook}{}
  \let\oldcaption\caption
  \renewcommand{\caption}[2][]{\oldcaption[#1]{\mycaptionhook{}#2}}

 \usepackage{fourier}

  \usepackage[version=3]{mhchem}

  \providecommand{\mysc}[1]{\MakeUppercase{#1}}
  \providecommand{\upsc}[1]{\MakeUppercase{#1}}

}

\usepackage{my_acronyms}

\usepackage[normalem]{ulem}

\usepackage[caption=false]{subfig}

\newcommand{\ie}{\emph{i.e.}}
\newcommand{\eg}{\emph{e.g.}}

\newcommand{\tl}{\text{l}}

\newcommand{\abs}[1]{\ensuremath{\lvert{#1}\rvert}}
\newcommand{\ket}[1]{\ensuremath{\lvert{#1}\rangle}}

\newcommand{\vect}[1]{\vec{#1}}

\renewcommand{\d}{\mathrm{d}}
\newcommand{\mat}[1]{\boldsymbol{#1}}
\newcommand{\op}[1]{\hat{#1}}

\newcommand{\csloc}{c_{s,\mathrm{loc}}}
\newcommand{\uloc}{u_{\mathrm{loc}}}
\newcommand{\cmatch}{c_{s,\mathrm{match}}}
\newcommand{\umatch}{u_{\mathrm{match}}}

\DeclareSymbolFont{greekletters}{OML}{zplm}{m}{it}

\usepackage[inline]{enumitem}
\setlist[enumerate]{label=\alph*)}%
\appto\mycaptionhook{%
  \setlist[enumerate]{label=\textbf{\alph*})}%
  \crefname{enumi}{panel}{panels}%
  \Crefname{enumi}{Panel}{Panels}%
}

\begin{document}

\title{Breaking a superfluid harmonic dam: Observation and theory of Riemann invariants and accelerating sonic horizons}

\author{Shashwat Sharan}
\affiliation{Department of Applied Mathematics, University of California, Merced, Merced, CA, USA 95343}

\author{Judith Gonzalez Sorribes}
\affiliation{Department of Physics and Biophysics, University of San Diego, San Diego, CA, USA 92110}

\author{Patrick Sprenger}
\affiliation{Department of Applied Mathematics, University of California, Merced, Merced, CA, USA 95343}

\author{Mark A. Hoefer}
\affiliation{Department of Applied Mathematics, University of Colorado, Boulder, CO, USA 80309}

\author{P.~Engels}
\affiliation{Department of Physics and Astronomy, Washington State University, Pullman, WA, USA 99164}

\author{Boaz Ilan}
\email{bilan@ucmerced.edu}
\affiliation{Department of Applied Mathematics, University of California, Merced, Merced, CA, USA 95343}

\author{M.~E.~Mossman}
\email{mmossman@sandiego.edu}
\affiliation{Department of Physics and Biophysics, University of San Diego, San Diego, CA, USA 92110}
\affiliation{Department of Physics and Astronomy, Washington State University, Pullman, WA, USA 99164}

\begin{abstract}
\noindent 

An experimental and theoretical study of sonic horizons emerging from the dam-break problem in a Bose-Einstein condensate confined in an anisotropic harmonic trap is presented. Measurements, analysis, and numerics reveal the formation of a sonic horizon that undergoes acceleration due to harmonic confinement. The superfluid is characterized using a robust measurement technique to determine Riemann invariants. Experimental observations agree with an analytical solution of the Gross-Pitaevskii equation and computations. The collision and annihilation between two sonic horizons at long times is predicted. 

\end{abstract}

\maketitle
\clearpage

The concept of an acoustic or sonic horizon (SH) and of a sonic black hole was first proposed in Ref.~\cite{unruh1981experimental}.
This spawned an emerging area of exploration of analogues of singular gravitational phenomena, cf.~\cite{garay2000sonic,visser2002artificial,Barcel:2003,Mayoral:2011,Horstmann:2011,wang2016sonic,Tsuda_2023,Fischer_2023}.
SHs have recently been observed in both classical and quantum systems, including water waves~\cite{rousseaux2008observation,Torres:2017,rozenman2024observation},
\mbox{Bose-Einstein condensates (BECs)~\cite{Lahav:2010,steinhauer2014observation,Steinhauer:2016,
MunozdeNova:2019,Kolobov:2021,Tamura_2023}},
and nonlinear optics~\cite{belgiorno2010hawking,drori2019observation,Falque_2025,vocke2018rotating}.
To date, experimentally reported SHs consist of a single, isolated horizon,
either static or moving at constant velocity relative to the laboratory frame. 
By contrast, an accelerating SH would serve as a closer analogue of an accelerating black hole, whose thermodynamics differ markedly from those of a static black hole~\cite{appels2016thermodynamics}.
Moreover, the possibility of interactions between two or more SHs, such as their collision and annihilation, arises naturally, cf.~\cite{barcelo2019analogue}, but has not yet been explored.

In this Letter, we observe an accelerating SH in a BEC and propose a framework for studying the interaction of \mbox{multiple} SHs.
To generate these horizons, we experimentally realize the classical dry-bed dam-break setup, in which a \mbox{large barrier} separates high density and vacuum regions.
Dam-break problems have been studied extensively in both classical and quantum \mbox{systems}, where they give rise to rich nonlinear dynamics such as \mbox{expansion} waves, shocks, and turbulence~\cite{Miles_1990,Bonnefoy_2020,von_H_fen_2022,Garoosi_2022,Marangoz_2024,Muchiri_2024}, and their analytical tractability have motivated applications in nonlinear optics~\cite{Xu_2017,Audo_2018,bienaime_quantitative_2021,Dieli_2024}, magnetohydrodynamics~\cite{Wu_1995,GIACOMAZZO_2006,Takahashi_2013,Minoshima_2021}, magneto-gas-dynamics~\cite{Singh_2014}, and BECs~\cite{Chang_2008,Mossman_2018,MossmanS_2024}.
In our setup, we employ a harmonically trapped channel-geometry BEC, where a repulsive optical barrier initially separates the BEC from an adjacent vacuum region. 
The barrier is removed \mbox{instantaneously}, triggering an expansion wave, also known as a rarefaction wave (RW), flowing into the vacuum.
Analogous to bathymetry in shallow-water waves, the harmonic confinement of the trap provides an inhomogeneous background landscape that proves critical for the acceleration of ensuing SHs.

To detect and study SH dynamics, we base our approach on Riemann invariants (RIs), which are central to the analysis of hyperbolic conservation laws ~\cite{lax_hyperbolic_1973,Leveque2002,whitham2011linear}.
RIs are combinations of hydrodynamic variables that, in the absence of external forcing, are constant along characteristic space-time curves.
If all but one of the RIs are constant, the system admits what is known as a \emph{simple wave}~\cite{Leveque2002}. 
Simple waves arise with piecewise constant initial data (Riemann data), where exact analytical solutions are known.
However, real-world systems depart from such idealized conditions due to inhomogeneous initial data, external forces, dissipation, and other physical mechanisms. 
These deviations disrupt the simple-wave picture and give rise to more complex dynamics. 
Nevertheless, the RIs and their associated characteristic velocities can be used to identify the location of SHs. 
RIs thus provide a natural framework for characterizing SHs.

The evolution of the BEC  is modeled by the three-dimensional (3D) Gross-Pitaevskii equation (GPE) with a harmonic potential~\cite{pitaevskii2016bose}.
Due to strong radial confinement by the trap, a reduced one-dimensional (1D) model captures the essential features of the flow. 
In the absence of quantum pressure, this 1D model is analogous to the shallow-water equations with bathymetry, for which simple waves do not exist.
Remarkably, we find exact analytical solutions of the shallow-water equations with parabolic bathymetry that also solve the 1D GPE with a harmonic potential. 
This solution generalizes RWs in the presence of harmonic confinement and helps provide an analytical characterization of the flow.

Determining the RIs experimentally requires measurement of the local density and flow velocity of the BEC.
To this end, we devise a barrier-pulse procedure that simultaneously measures the local flow velocity and speed of sound, from which the RIs are inferred directly. 
The observed RI values are in agreement with 1D analytical solutions and computations of the 3D GPE.
For short times, a single SH exists. 
As the system evolves, the interaction with the harmonic potential induces an acceleration of the SH. 
At later times, our theory predicts that two additional SHs form and persist simultaneously with the first one. 
Moreover, we predict that two SHs collide and annihilate under properly timed conditions.
This could provide a future avenue for studying analogues of accelerating and interacting black and white holes.

To provide context for the discussion of the superfluid dam break, we begin with a brief discussion of the experimental setup. See End Matter for more details.
A BEC composed of \ce{^{87}Rb} atoms is held in a cigar-shaped optical dipole trap with tight confinement in the $y$ and $z$ directions. 
To create an effective dam, a blue-detuned laser  propagating in the \mbox{$z$-direction} is employed that can be shifted along the \mbox{$x$-axis}.
The potential produced by this laser in the $y$-direction extends well beyond the tight radial confinement of the cloud, so that the atoms do not tunnel through the barrier when traveling at the speeds considered in this study.

\begin{figure}[t!]
    \centering
    \includegraphics[width=0.95\columnwidth]{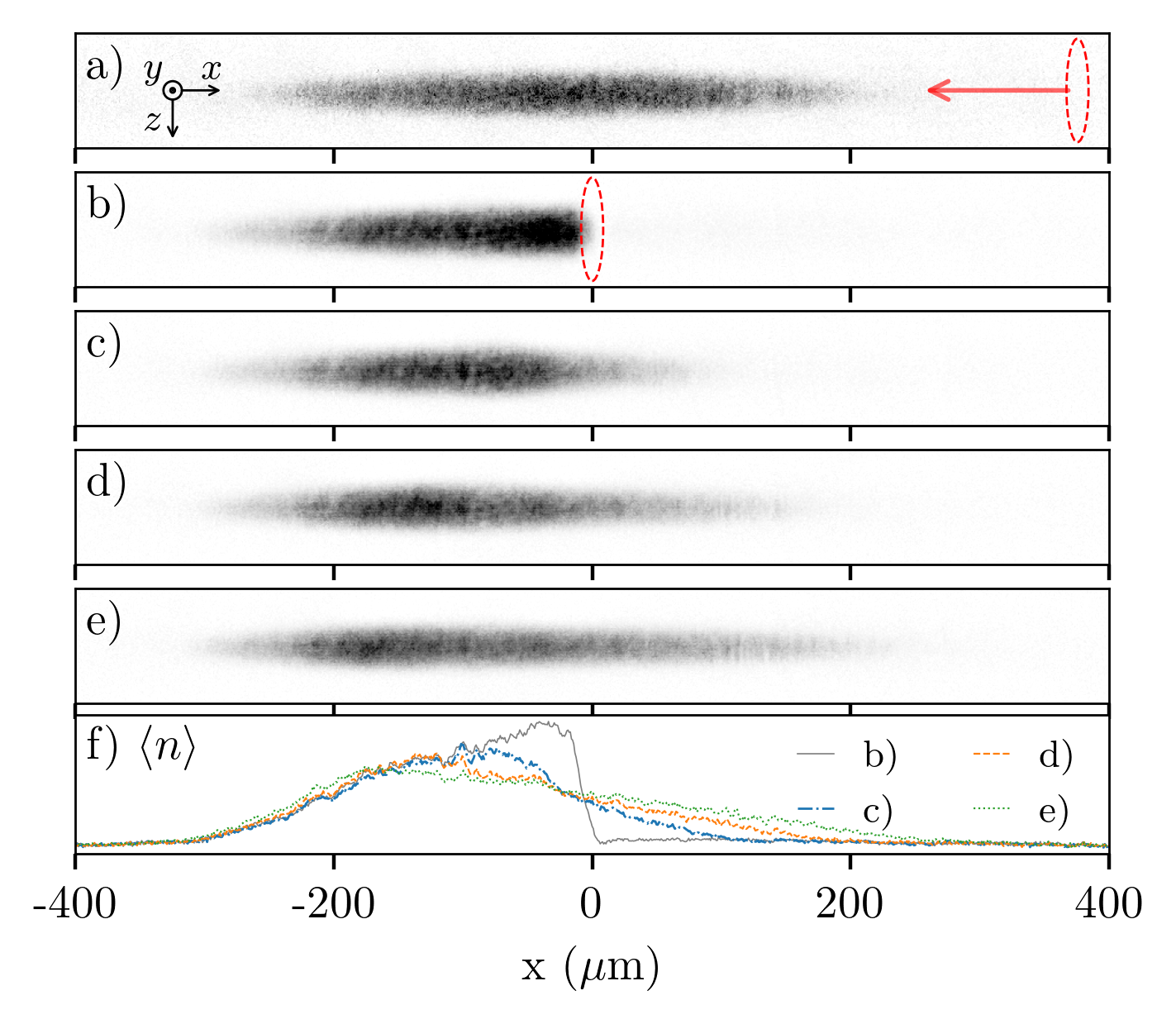}
    \caption{Evolution of a rarefaction flow. 
    a) A repulsive optical barrier or dam (represented by the red dashed oval) is slowly swept 523(1)~\textmu{m} from the right to the center of the BEC in 2~s. 
    b) The dam is removed at $t=0$ and atoms are allowed to flow into the $x>0$ region of the trap, \eg~for  
    c) $t=20$~ms, d) $t=40$~ms, and e) $t=60$~ms. 
    f) Integrated cross sections of the density are provided for panels b)-e), where the canonical parabolic profile appears at short times. 
    Data has been averaged over five experimental runs of the same parameters.}
    \label{fig:rareedge}
\end{figure}

To generate a controlled rarefaction flow, we use this dam potential to create a dry-bed dam-break scenario. This procedure is depicted in the experimental images of Fig.~\ref{fig:rareedge}\mbox{a--e}.
After an adiabatic sweep of the potential from the outside (Fig.~\ref{fig:rareedge}a), the BEC is in a Thomas-Fermi (TF) ground state, occupying the left half ($x<0$) of the harmonic trap (Fig.~\ref{fig:rareedge}b).
When the dam potential is suddenly removed, the BEC rarefies into the previously unoccupied region  ($x>0$), as shown in Fig.~\ref{fig:rareedge}\mbox{c--e}.
The observed density profiles are highlighted in Fig.~\ref{fig:rareedge}\mbox{f}. 
The rarefaction flow in our experimental setting extends over \mbox{hundreds} of microns, providing ample space for the propagation and investigation of hydrodynamic features.

We characterize the rarefaction flow in our system by determining the RIs. To facilitate this analysis, we model the flow using an effective 1D GPE~\cite{kevrekidis2015defocusing},
\begin{align}\label{eq:1DGP}
    i \hbar \frac{\partial  \psi}{\partial t} = \bigg(-\frac{\hbar^2}{2m} \frac{\partial^2}{\partial x^2} + \frac{1}{2}m \omega_x^2 x^2 + g_{1\mathrm{d}} |\psi|^2\bigg) \psi,
\end{align}
where $\hbar$ is the reduced Planck constant, $m$ is the atomic mass of \ce{^{87}Rb}, and $g_{1\mathrm{d}}$ is the effective 1D interatomic coupling constant. This reduction is valid when the transverse confinement is significantly stronger than the axial confinement, and the system can be treated as effectively 1D. \mbox{In this model, $g_{\rm 1d} = k_\mathrm{B}\times0.0276$~nK \textmu m (see End Matter)}, and $\psi$ is normalized such that $\int |\psi|^2 \,{\rm d}x = \mathcal{N}$ is the total number of atoms in the BEC.

The 1D GPE~\eqref{eq:1DGP} can be reformulated in terms of \mbox{hydrodynamic variables }using the Madelung transformation, $\psi(x,t) = \sqrt{n} e^{im\phi/\hbar}$, where $n(x,t) = |\psi|^2$ denotes the local atom density and $\phi(x,t)$ is the velocity potential, so the flow velocity in the $x$ direction is $u(x,t) = \partial\phi/\partial x$. 
Substituting this transformation into Eq.~\eqref{eq:1DGP} yields the system
\begin{subequations}
\label{eq:hydro_sys_with_dispersion}
\begin{eqnarray}
\label{eq:mass_consevation}
\frac{\partial n}{\partial t} + \frac{\partial (n u)}{\partial x} &=& 0,\\
\frac{\partial (nu) }{\partial t}
+ \frac{\partial}{\partial x}\Big( n u^{2} + \frac{g_{1\mathrm{d}} n^{2}}{2m} 
- \frac{\hbar^{2}n}{4m^{2}} \frac{\partial^2 \log n}{\partial x^2} \Big) 
&=& -n \omega _{x}^{2} x. \hspace{0.15in}
\label{eq:momentum_consevation}
\end{eqnarray}
\end{subequations}
Equation~\eqref{eq:mass_consevation} describes local mass conservation while Eq.~\eqref{eq:momentum_consevation} describes momentum balance, which includes hydrodynamic pressure ($\propto g_\mathrm{1d}$), quantum pressure ($\propto \hbar^2$), and the harmonic trap ($\propto \omega_x^2$).
Since our system is in the TF regime, the quantum pressure is negligible compared to the hydrodynamic pressure (see End Matter). Ignoring this term for the moment allows the system to be diagonalized in terms of the RI variables,
\begin{align}
\label{eq:riemann_inv}
r_\pm = \frac{u}{2} \pm c_{s}, \quad c_s = \sqrt{\frac{g_{1\mathrm{d}} n}{m}},
\end{align}
where $c_{s}$ is the local speed of sound. 
We refer to  $r_+$  as the fast RI and  $r_-$  as the slow RI because their corresponding characteristic speeds,  $v_{\pm} = u \pm c_s$, satisfy  $v_+ > v_-$, making  $r_+$  associated with faster wave propagation. 
The diagonalization results in the following system for $r_\pm$,
\begin{equation}\label{eq:dispersionless_diagonal_system}
 \frac{\partial r_{\pm}}{\partial t} 
+ v_{\pm} \frac{\partial r_{\pm}}{\partial x} 
= -\frac{1}{2} \omega_{x}^{2} x, \quad v_{\pm} =\frac{1}{2}(3r_{\pm}+r_{\mp}).
\end{equation}

In the absence of the trap ($\omega_x=0$), this diagonal system admits the self-similar solution
\begin{equation}\label{eq:slow_RW}
    r_{+}(x,t) = s_0, \quad r_{-}(x,t) = -\frac{s_0}{3} + \frac{2x}{3t},
\end{equation}
which describes the expanding region of a slow, simple RW.
Here,  $s_0=\sqrt{\mu/g_{1\mathrm{d}}}$ is the local speed of sound at the initial peak density, where $\mu$ is the chemical potential of the TF ground state. This solution is classified as simple because the fast RI, $r_+$, is constant while the slow RI exhibits self-similar dependence on $x/t$ within the expansion region. The solution~\eqref{eq:slow_RW} is defined within the rarefaction region $-s_0 t < x < 2 s_0 t$.
The right edge is a vacuum point, where the density vanishes.

A SH occurs when either $v_{+}=0$ or $v_{-} = 0$. 
The simple-wave solution \eqref{eq:slow_RW} possesses a single stationary SH at $x=0$ corresponding to $v_- =0$.
When a harmonic potential is present, we show by finding an exact analytical solution that inside the rarefaction region multiple SHs can coexist and interact.
We present an exact solution of Eq.~\eqref{eq:dispersionless_diagonal_system},
\begin{subequations}
\label{eq:rp_rm_exact_solution}
\begin{eqnarray}
       r_{+}(x,t) &=&  \frac{s_0\cos\big(\tfrac{\omega_x t}{4}\big) - \tfrac{\omega_x x}{2}\sin\big(\tfrac{3\omega_x t}{4}\big)}{\cos\big(\tfrac{3\omega_x t}{4}\big)},\\
        r_{-}(x,t) &=& \frac{-s_0 \sin\big(\tfrac{\omega_x t}{4}\big) + \tfrac{\omega_x x}{2} \cos\big(\tfrac{3\omega_x t}{4}\big) }{\sin\big(\tfrac{3\omega_x t}{4}\big)}.
\end{eqnarray}
\end{subequations}
In the vanishing potential limit ($\omega_x \to 0$), solution~\eqref{eq:rp_rm_exact_solution} reduces to the simple RW solution~\eqref{eq:slow_RW}, indicating that at short times $r_+$ remains approximately conserved while $r_{-}$ displays a nearly self-similar profile in $x/t$.
At longer times, this solution possesses a temporal singularity at $t = \tfrac{T_{\rm har}}{3}$, where \mbox{$T_{\rm har} = \frac{2\pi}{\omega_x}$} is the harmonic trap period. The presence of this singularity corresponds physically to the occurrence of steep gradients in the hydrodynamic variables.
On a related note, this solution may also have relevance for shallow-water waves propagating over a parabolic bathymetry, which have recently been studied analytically~\cite{Camassa_2020,camassa2022evolution}. 

The expressions for the local speed of sound and the flow velocity can be obtained using Eq.~\eqref{eq:rp_rm_exact_solution} as $ c_s = \tfrac{1}{2} (r_{+} - r_{-})$ and $u = r_{+} + r_{-}$, respectively, yielding 
\begin{subequations} 
\label{eq:density_phase_exact_solution} 
\begin{eqnarray} 
c_{s}(x,t) &=& \frac{\omega_x}{2}\csc\Big(\frac{3\omega_x t}{2}\Big) \big(x_r(t) - x \big),\\
u(x,t) &=& \frac{2s_0 + \omega_x x \cos\big(\frac{3\omega_x t}{2}\big) \csc\big(\frac{\omega_x t}{2}\big)}{1+2\cos(\omega_x t)}~,
\end{eqnarray} 
\end{subequations}
where the vacuum point $x_r(t) = \frac{2s_0}{\omega_x} \sin(\omega_x t)$ defines the right edge of the rarefaction flow. 
Importantly, the quantum pressure evaluated for this solution vanishes and thus Eq.~\eqref{eq:density_phase_exact_solution} satisfies the full 1D hydrodynamic system~\eqref{eq:hydro_sys_with_dispersion} and, consequently, the 1D GPE~\eqref{eq:1DGP} with a harmonic potential.

Direct simulations of the 1D GPE show that, at short times, the BEC dynamics are described well by solution~\eqref{eq:density_phase_exact_solution} within the rarefaction region $x_l(t)<x<x_r(t)$. 
The left edge of this region is given by (see End Matter)
\begin{equation}
    \label{eq:xl}
    x_l(t) = -\frac{\sqrt{2} s_0}{\omega_x}\sin\left(\frac{\omega_x t}{\sqrt{2}}\right)~.
\end{equation}
Near the vacuum point, numerical simulations show that the density profile smoothly connects to the vacuum region, consistent with the quadratic density dependence $c_s^2 \propto (x_r(t)-x)^2$. Consequently, Eq.~\eqref{eq:density_phase_exact_solution} accurately captures the dynamics near the vacuum point up until the singularity time.
As time progresses, however, substantial deviations from the analytical solution emerge, particularly at locations away from the vacuum point.

To accurately capture the BEC dynamics away from the vacuum point, we construct a \emph{matched} solution within the rarefaction region. We assume that the true speed of sound and flow velocity are well-approximated by spatially quadratic profiles with time-dependent coefficients
\begin{subequations}
\label{eq:matched_solutions}
\begin{eqnarray}
    \cmatch(x,t) &=& c_0(t) + c_1(t) \, x + c_2(t) \, x^2,\\
    \umatch(x,t) &=& u_0(t) + u_1(t) \, x + u_2(t) \, x^2.
\end{eqnarray}
\end{subequations}
The coefficients $c_j(t)$ and $u_j(t)$ for $j \in \{0,1,2\}$ are determined by requiring that the matched profiles (i) connect smoothly with solution~\eqref{eq:density_phase_exact_solution} at $x=x_r(t)$, and (ii) remain continuous with the initial stationary TF profile at $x=x_l(t)$ (coefficients provided in End Matter). 
Using this matched solution, the RIs and SHs inside the rarefaction region are discussed in the context of the experimental results below.

\begin{figure}
    \centering
    \includegraphics[width=0.95\columnwidth]{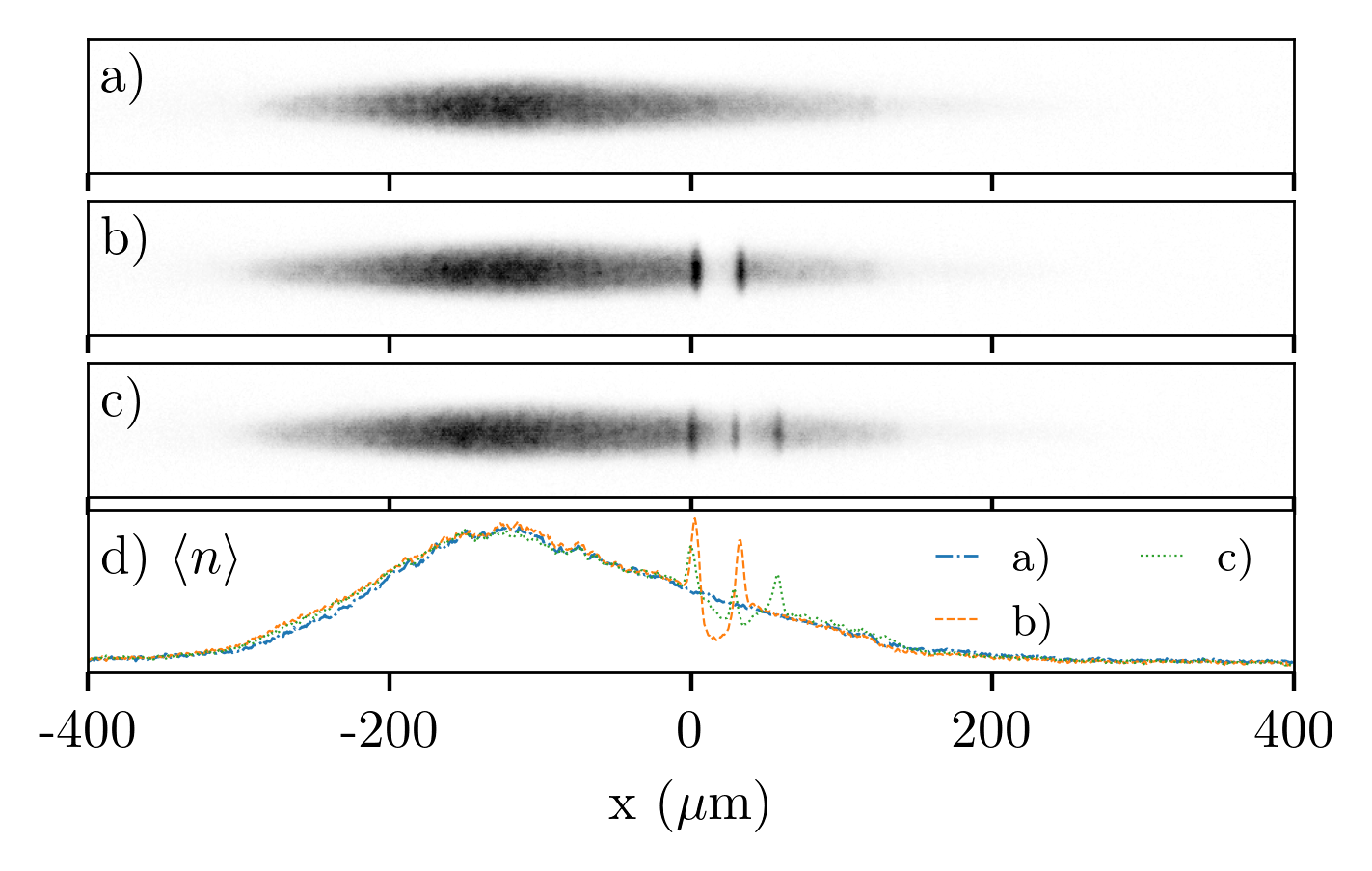}
    \caption{Barrier pulse procedure. 
    a) A rarefaction flow imaged at $t=40~$ms prior to pulsing. 
    b) The barrier is introduced around $x_c=21$~\textmu{m}, pulsed for 0.5~ms, and imaged. 
    c) Excitations are imaged 5~ms after the pulsing. 
    d) Integrated cross sections of the density from panels (a)-(c) with arbitrary units. 
    All images taken in a 5~ms time-of-flight. Images are averaged over 10 experimental runs with the same parameters.
    }
    \label{fig:expt_procedure}
\end{figure}

The central role played by the RIs motivates the development of an experimental method to robustly measure these quantities.
Here, we describe such a method using a barrier-pulse procedure.
For this, first a rarefaction flow is generated, as described in Fig.~\ref{fig:rareedge}.
We then pulse the barrier potential on for 0.5~\rm{ms} at a new position $x_c$. 
This generates two sharp, high-density excitations on either side of the barrier.
Relative to the rarefaction flow, these excitations travel outwards at approximately the speed of sound~\cite{Andrews_1997,Andrews_1998,Steinhauer_2002,Chang_2008}. 
This experimental procedure is demonstrated in Fig.~\ref{fig:expt_procedure}.
\mbox{To determine the RIs,} we first characterize the local flow velocity and speed of sound in the system by measuring the positions of the two high-density excitations at $0$ and $5$~ms after the barrier pulse.
Finding the change in position of the excitations over a known change in time allows us to experimentally extract the fast- and slow-characteristic velocities $v_\pm$ of the high-density peaks.
The local background flow velocity is then obtained by averaging these velocities, $\uloc=\tfrac{1}{2}(v_{+}+v_{-})$, while the local speed of sound is found by subtracting the local flow velocity from the right-moving velocity, $\csloc = v_{+}-\uloc=\tfrac{1}{2}(v_{+}-v_{-})$.
From these local hydrodynamic variables, the two measured RIs are computed using Eq.~\eqref{eq:riemann_inv}. 
By repeating this procedure for various positions $x_c$ and rarefaction times $t$, we experimentally determine the spatio-temporal dependence of the RIs.

\begin{figure}
    \centering
    \includegraphics[width=0.95\columnwidth]{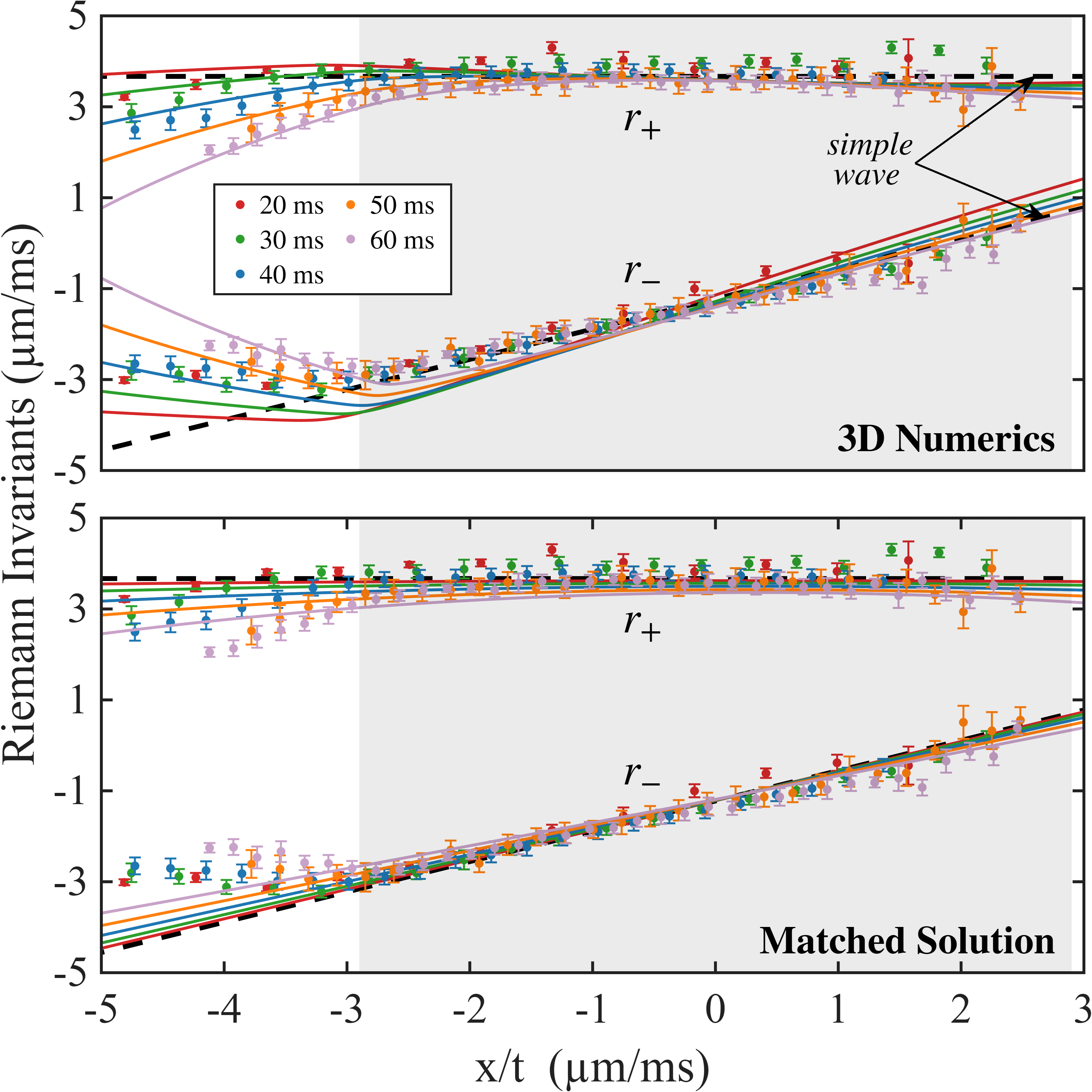}
    \caption{Experimentally measured slow ($r_{-}$) and fast ($r_{+}$) RIs along the channel, plotted against the similarity variable $x/t$ at different rarefaction flow times for $t\le 60$ ms (data points with error bars, see legend).
    Error bars of the measurement represent the standard deviation of the mean at each probed location. 
    The 3D GPE numerical simulations (solid curves, upper panel) and matched solutions from Eq.~\eqref{eq:matched_solutions} (solid curves, lower panel) are shown for comparison. 
    The shaded region indicates where the matched solution is valid. 
    }
    \label{fig:hydrodata}
\end{figure}

Figure~\ref{fig:hydrodata} presents experimental and theoretical values of \(r_{\pm}\) at five rarefaction times (\(t \leq 60\) ms) and at various spatial locations. To facilitate a direct comparison of datasets \mbox{at different times}, the values are plotted against \(x_c/t\) for various pulse locations \(x_c\). When scaled this way, Fig.~\ref{fig:hydrodata} shows that, 
initially, within the rarefaction region ($x>0$), 
the experimental values align closely with the universal simple slow RW solution~\eqref{eq:slow_RW}. 
In particular, the fast RI is constant, $r_+=s_0\approx3.67$~\textmu{m}/ms, while the slow RI, $r_-$, has a \mbox{$2/3$ slope (black dashed lines)}.
This demonstrates the initial self-similarity.
However, the slopes of $r_\pm$ 
decrease over time due to the harmonic confinement
in a manner consistent with the exact solution~\eqref{eq:rp_rm_exact_solution}. 
We note that the match between experimental data and theory is not as good for $r_-$ than $r_+$, especially for $x<0$, as any imprecision in the measurement compounds when calculating $r_-$ while they cancel for $r_+$.

\begin{figure}
    \centering
    \includegraphics[width=0.99\columnwidth]{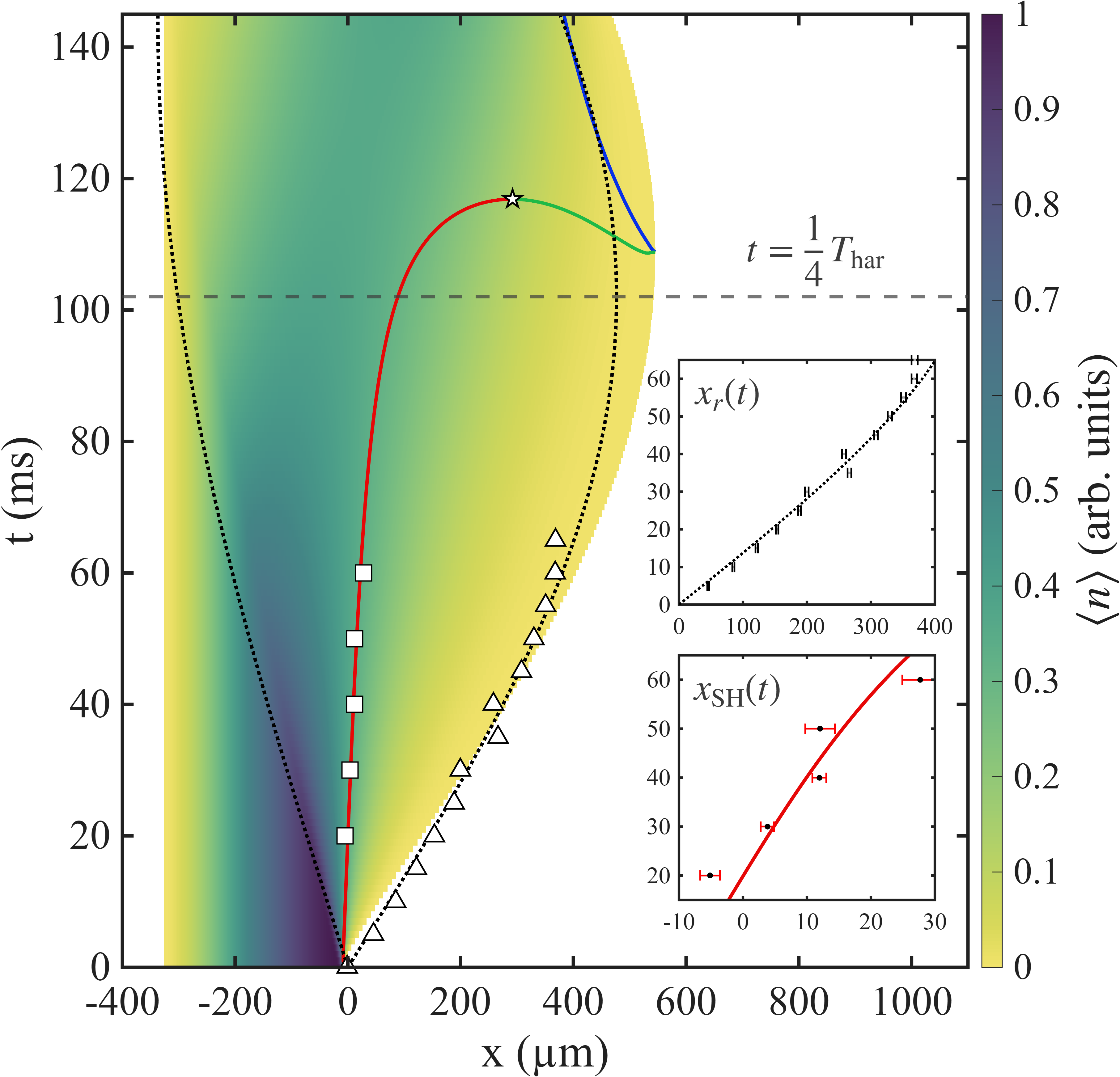}
    \caption{Evolution of the condensate density and sonic horizons. 
    Background (see color bar) is the integrated cross-sectional density $\langle n \rangle$ from the 3D GPE simulations.
    Solid curves track three SHs: two ``slow'' horizons (red and green, where $u_{\rm loc} = c_{s,\rm loc} $) with their collision marked by a star, and one ``fast'' horizon (blue, where $u_{\rm loc} = -c_{s,\rm loc}$).
    Dotted black curves are predictions for the rarefaction edges from Eqs.~\eqref{eq:density_phase_exact_solution}-\eqref{eq:xl}.
    Experimental data for the SH (squares) and right edge of the rarefaction (triangles) are overlaid. 
    The horizontal dashed line indicates the 1D turnaround time, $\frac{1}{4}T_{\rm har}$.
    Insets magnify the vacuum point ($x_r(t)$) and SH locations ($x_\mathrm{SH}(t)$) for $t \leq 60$~ms, with error bars showing the standard deviation of the mean from multiple experimental runs.
    }
    \label{fig:fastR_WH}
\end{figure}

We now leverage the data obtained from these experiments, simulations of the 3D GPE (see End Matter for details) to describe the dynamics of SHs that arise in harmonically confined rarefaction flows, as shown in Fig.~\ref{fig:fastR_WH}.
A first SH emanates from the origin (red solid curve).
Here, excitations are unable to move through the SH, realizing an acoustic white hole~\cite{Mayoral:2011,demirkaya2019analog,yang2019sonic}.
As the rarefaction flow evolves, the SH accelerates away from $x=0$ in the positive direction, \mbox{as shown in the lower inset of Fig.~\ref{fig:fastR_WH}.}

Figure~\ref{fig:fastR_WH} also shows that one slow SH (green curve) and one fast SH (blue curve) emerge from the vacuum point at around the turnaround time, 
$t\approx108~\mathrm{ms}$,
when the edge of the flow momentarily comes to rest and reverses direction. 
This is somewhat greater than the turnaround time predicted by the analytical solution~\eqref{eq:rp_rm_exact_solution},
\ie,
$\tfrac{1}{4}T_\mathrm{har} \approx 102$~ms.
We attribute this difference to transverse BEC dynamics that the 1D GPE does not capture.
The two slow horizons (red and green curves) collide and annihilate at $t \approx 117$~ms (marked by a star). To the best of our knowledge, this is the first prediction of annihilation of sonic horizons.
An experimental verification for these dynamics presents an intriguing outlook for future work. 

In summary, we presented a robust experimental method for determining the RIs in a BEC and implemented it in a dry-bed dam-break of a BEC confined in a harmonic trap.
This method can be applied to other cold-atom experiments to determine local densities, velocities, and RIs. 
We derived an exact analytical solution of the 1D evolution of the BEC in the presence of a harmonic trap, which agrees well with the experiments and numerical simulations of the 3D GPE. 
Our experimental and theoretical results demonstrate the formation and subsequent acceleration of a SH.  
Our theory predicts the emergence of two additional SHs, and the collision and annihilation of two SHs.
The analytical solution of the 1D GPE (and shallow-water equations) with harmonic confinement could inform other studies in classical and quantum systems, such as dam-break reflections~\cite{Hogg_2021,Ungarish_2022} and interface singularities~\cite{camassa2022evolution}.

Accelerating black holes have recently attracted much attention due to their thermodynamics~\cite{appels2016thermodynamics,anabalon2018holographic,anabalon2019thermodynamics} and the possibility of observing them ~\cite{ashoorioon2022distinguishing}.
This work introduces a line of questions for future studies, such as the thermodynamics of accelerating sonic analogues of black and white holes, and the long-time dynamics of SHs in non-uniform background landscapes.

\begin{acknowledgments}
J.G.S.\ and M.E.M.\ acknowledge support from the \gls{NSF} through Grant No.\ \mysc{PHY-2137848}. M.E.M additionally acknowledges funding from \gls{NSF} through Grant No.\ \mysc{DMS-1941489} and from the Clare Boothe Luce Professorship Program of the Henry Luce Foundation.
P.E.\ acknowledges support from the \gls{NSF} through Grant No.\ \mysc{PHY-2207588} and from a Boeing Endowed Professorship at WSU.
M.A.H.\ acknowledges support from \gls{NSF} via Grant No.\ \mysc{DMS-2306319}.
P.E., M.A.H., S.S., P.S., and B.I. would like to thank the Isaac Newton Institute for Mathematical Sciences, Cambridge, for support and hospitality during the programme "Emergent phenomena in nonlinear dispersive waves", where work on this paper was undertaken. This programme was supported by EPSRC grant EP/R014604/1.
\end{acknowledgments}

\paragraph{Author Contributions}
J.G.S., P.E., and M.E.M. conceived and performed the experiments and data analysis.
S.S., P.S., B.I., and M.A.H. performed theoretical calculations and numerical simulations.
All authors discussed the results and contributed to the writing of the manuscript.

\paragraph{Competing Interests}
The authors declare no competing interests.

\paragraph{Materials \& Correspondence}
Please direct any questions or requests concerning this article to M.~Mossman or B.~Ilan.

\paragraph{Code Availability}
All relevant code used for numerical studies in this work is available from the corresponding authors upon reasonable request.

\paragraph{Data Availability}
All relevant experimental and numerical data sets in this work will be made available from the corresponding authors upon reasonable request.

%

\section{End Matter}
\label{sec:supp}

\subsection*{Experimental procedure}

\begin{figure}[h]
     \centering
     \includegraphics[width=2.25in]{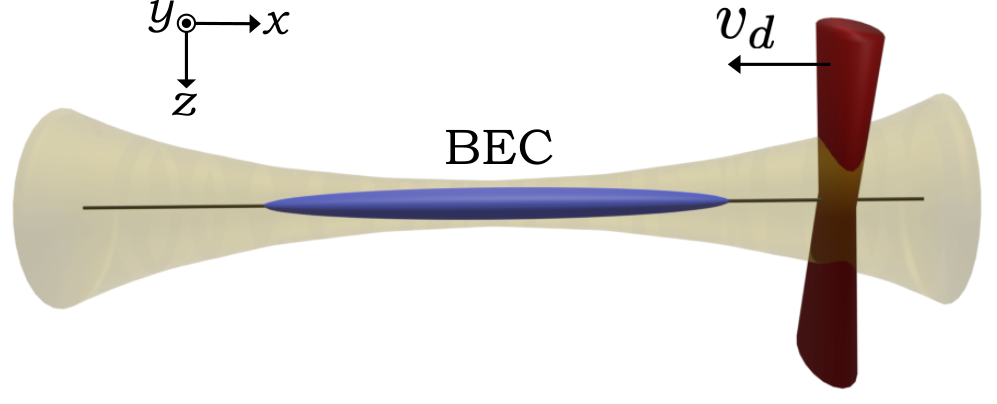}
     \caption{General experimental setup. An elongated BEC (blue) trapped in an optical dipole trap (yellow). A repulsive barrier (red) can move along the $x$-axis, along the BEC.}
     \label{supp:setup}
 \end{figure}

The general setup of our experiment consists of a BEC of $1.1\times10^6$ $^{87}$Rb atoms condensed into the in the \mbox{$\ket{F,m_F}=\ket{1,-1}$} state (see Fig.~\ref{supp:setup}). The BEC is confined in an optical dipole trap generated by a 1064~nm laser beam propagating horizontally along the $x$-axis. This beam is tightly focused to a 20~\textmu{m} waist, producing harmonic confinement in the vicinity of the focus characterized by harmonic trap frequencies of $\{\omega_x,\omega_y,\omega_z\}=2\pi\times\{2.45,243,251\}$~Hz. Here, $y$ denotes the imaging direction and $z$ is aligned with gravity. The atom numbers and potential result in a chemical potential of $k_\mathrm{B} \times 104$~nK in the trap.

To manipulate the BEC, we employ a repulsive laser beam generated by a 660~nm laser, directed along the 
$z$-axis, as shown in Fig.~\ref{supp:setup}. 
This beam has a width of 54.5~\textmu{m} along the $y$-axis, significantly larger than the BEC width of approximately 5~\textmu{m}, effectively forming a uniform light sheet at the location of the atoms. The beam intensity is set such that the resulting repulsive potential exceeds the BEC’s chemical potential by a factor of four. The beam is translated along the $x$-axis with the help of a galvanometer. 

Individual procedures for the presented experiments are described in the main text. For example, to create a reproducible rarefaction flow, we move the barrier from the right outside of the BEC to the center of the BEC at $v_d=0.261(1)$~\textmu{m}/ms, well below the speed of sound in the BEC. The potential is then turned off for a fixed amount of time, allowing the system to rarefy into the empty section of the dipole trap.
All imaging is performed after 5~ms time-of-flight, during which all optical potentials are switched off and the BEC is freely expanding.

\subsection*{Theoretical model and Computational methods}

The propagation of the BEC is modeled by the 3D GPE,
\begin{equation}\label{supeq:3dGP}
i\hbar \, \frac{\partial \Psi}{\partial t} = -\frac{\hbar^2}{2m} \nabla^2 \Psi + V(\mathbf{x}) \Psi + g_{3\rm d}|\Psi|^2 \Psi,
\end{equation}
where $m$ is the atomic mass of $^{87}\text{Rb}$,
$V(\mathbf{x})$ is an external potential,
$g_{3\rm d} = \frac{4\pi\hbar^2 a_s}{m}$ is the interatomic coupling coefficient determined by the scattering length 
$a_s = 100.4~a_0$, where $a_0$ is the Bohr radius.
The wavefunction $\Psi(\mathbf{x},t)$ is normalized such that 
$\int |\Psi|^2 \, d\mathbf{x} = \mathcal{N}$ is the total number of atoms.
For $t>0$, the only potential is harmonic with \mbox{$V(\mathbf{x}) = \frac{1}{2} m (\omega_x^2 x^2 + \omega_y^2 y^2 + \omega_z^2 z^2)$}.
\
The initial wavefunction $\Psi({\bf x},0)$ is 
the ground state obtained with the harmonic trap and the dam potential 
\begin{equation}\label{supp:3d_dam_potential}
V_{\rm d}(\mathbf{x}) = U_d H(-x) \exp{\Big(-\frac{2x^2}{s_x^2} 
- \frac{2y^2}{s_y^2}\Big)} + U_d H(x),
\end{equation}
where $H(x)$ is the Heaviside step function and
the peak height of this potential is $U_d = k_\mathrm{B}\times408$~nK, such that the atoms do not tunnel through the barrier. 

The computational domain is a rectangular box with lengths {$\{L_x, L_y,L_z\} = \{900, 5, 5\}$~\textmu{m}}, which is discretized with {$\{N_x, N_y, N_z\} = \{2^{13}, 2^{5}, 2^{5}\}$} grid points.
The ground state is computed using a Newton-conjugate-gradient method \cite{yang_newton-conjugate-gradient_2009}, 
initialized with a chemical potential of 
$\mu= k_\mathrm{B}\times 140.65~\mathrm{nK}$, which reproduces the number of atoms in the experiment.

Time evolution is performed using a second-order pseudo-spectral split-step scheme~\cite{yang2010nonlinear} with a time step of \mbox{$\Delta t = 0.005$~ms}. The ground state serves as the initial condition and the local density and velocity are recovered via
\begin{equation}
n= |\Psi|^2, \qquad 
    \mathbf{u} = \frac{i\hbar}{2mn} ( \Psi \nabla \Psi^*-\Psi^* \nabla \Psi)~.
\end{equation}

The RIs are recovered from (see, e.g. Eq.~\eqref{eq:riemann_inv})
\begin{equation}
    r_{\pm} = \frac{\langle \mathbf{u} \cdot \hat{\mathbf{x}}\rangle}{2} \pm \sqrt{\frac{g_{1\rm d} \langle n \rangle}{m}},
\end{equation}
where $\mathbf{u} \cdot \hat{\mathbf{x}}$ is the component of the flow velocity along the axis of the BEC and the notation $\langle \cdot \rangle = \int \cdot \ \, \mathrm{d}y \, \mathrm{d}z$ denotes an integrated cross-section.


\subsection*{Negligible quantum pressure}
The dimensional 3D GPE Eq.~\eqref{supeq:3dGP} can be recast in nondimensional form using the rescaled coordinates
\begin{equation}
    t^{\prime} = \omega_z t, \quad \mathbf{x}^{\prime} = \frac{\mathbf{x}}{\ell}, \quad \Psi^{\prime} = \frac{\ell^{3/2}}{\sqrt{\mathcal{N}}}\Psi, \quad \ell = \left(\frac{4\pi \hbar^{2} a_{s}\mathcal{N}}{m^{2}\omega_{z}^{2}}\right)^{1/5}.
\end{equation}
This yields the nondimensional GPE
\begin{equation}\label{supeq:non_dim_3DGP}
i\varepsilon \frac{\partial \Psi}{\partial t} = -\frac{\varepsilon^{2}}{2} \nabla^{2} {\Psi} + \frac{1}{2} \big( \alpha_x {x}^{2} + \alpha_y {y}^{2} + \alpha_z{z}^{2} \big) {\Psi} + |{\Psi}|^{2} {\Psi},
\end{equation}
where the primes have been dropped. 
Here, $\alpha_j = (\omega_j / \omega_z)^2$ for $j \in \{x,y,z\}$ quantify relative strengths of the harmonic confinement in each direction. 
The dimensionless parameter
\begin{equation}
    \varepsilon = \left(\frac{\hbar}{m \omega_z (4\pi a_s \mathcal{N})^{2}}\right)^{1/5}
\end{equation}
quantifies the strength of quantum pressure. The wavefunction is normalized such that $\int |\Psi|^2 \, d\mathbf{x} = 1$. For parameters from experiment, we find $\varepsilon^2 = 9.15 \times 10^{-5} \ll 1$, \mbox{$\alpha_x = 9.52 \times 10^{-5} \ll 1$}, $\alpha_y = 0.94 \approx 1$ and $\alpha_z=1$.

Rewriting Eq.~\eqref{supeq:non_dim_3DGP} in hydrodynamic form via the Madelung transformation $\Psi = \sqrt{n} e^{i \Phi}$, where $\mathbf{u} = \nabla\Phi$, gives
\begin{equation}
    \begin{aligned}
\frac{\partial n}{\partial t} + \nabla \cdot (n \mathbf{u}) &= 0, \\
\frac{\partial (n\mathbf{u})}{\partial t} + \nabla \cdot (n\mathbf{u} \otimes \mathbf{u} + p_{\rm quantum}) &= -\nabla p_{\rm hydro} - n \nabla V,
\end{aligned}
\end{equation}
where the hydrodynamic and quantum pressures are given by
\begin{equation}
p_{\rm hydro} = \frac{1}{2} n^2, \quad p_{\rm quantum} = -\frac{\varepsilon^2}{4} n (\nabla \otimes \nabla) \log n.    
\end{equation}
This shows the quantum pressure, proportional to $\varepsilon^2$, away from sharp density transitions, is negligible compared to the hydrodynamic pressure, justifying its omission in the analysis.

\subsection{Interatomic coupling coefficient}

Neglecting quantum pressure, the interatomic coupling coefficient $g_{1\rm d}$ can be obtained from the (dimensional 1D) TF density profile
\begin{equation}\label{supp:TF_profile}
    |\psi_{\rm TF}| = \sqrt{\frac{\mu - V_{\rm h}(x) - V_{\rm d}(x)}{g_{1\rm d}}},
\end{equation}
which is valid in the TF region \(\Omega_{\rm TF} := \{\, x \in \mathbb{R} \,\, \big|\,\, \mu > V_{\rm h}(x) + V_{\rm d}(x) \,\}\).
Here, $V_{\rm h}(x) = \tfrac{1}{2} m \omega_x^2 x^2$ is the 1D harmonic potential, and the 1D dam potential is modeled as (see Eq.~\eqref{supp:3d_dam_potential})
\begin{equation}
    V_{\rm d}(x) = U_d H(-x)\,\exp\!\Big(-\frac{2x^2}{s_x^2}\Big) + U_d H(x).
\end{equation}

The normalization condition for $\psi_{\rm TF}$ is 
\(\int |\psi_{\rm TF}(x)|^2\,\mathrm{d}x = \mathcal{N}\).
Substituting \eqref{supp:TF_profile} into this condition gives
\begin{equation}\label{eq:g1d_integral_expression}
    g_{1\rm d}
    = \frac{1}{\mathcal{N}}
      \int_{\Omega_{\rm TF}}
      \Bigl(\mu - V_{\rm h}(x) - V_{\rm d}(x)\Bigr)\,\mathrm{d}x.
\end{equation}

For the present dam potential, $\Omega_{\rm TF}$ is well approximated by the interval \((-R_{\rm TF},-\,\tfrac{s_x}{\sqrt{2}})\),
with the upper limit valid when $m\omega_x^2 s_x^2 / U_d \ll 1$ and $\mu \ll U_d$. The lower limit is the TF radius,
\begin{equation}
  R_{\rm TF} = \sqrt{\frac{2\mu}{m\omega_x^2}}.
\end{equation}

Evaluating Eq.~\eqref{eq:g1d_integral_expression} under the assumption $s_x \ll R_{\rm TF}$ gives the analytic approximation
\begin{equation}\label{eq:g1d_analytical_expression}
    g_{1\rm d} = \frac{1}{\mathcal{N}} \left( \frac{2}{3}\mu R_{\rm TF} - \sqrt{\frac{\pi}{8}} U_d \, s_x\, \mathrm{erfc}(1)\right),
\end{equation}
where $\mathrm{erfc}(x)=1-\mathrm{erf}(x)$. For the experimental parameters, Eq.~\eqref{eq:g1d_analytical_expression} gives $g_{\rm 1d} = k_\mathrm{B}\times0.0286$~nK \textmu m.
The exact value of $g_{1\rm d}$, obtained by numerically evaluating Eq.~\eqref{eq:g1d_integral_expression} over the exact TF region, is 
$k_\mathrm{B}\times0.0276$~nK \textmu m. This shows that Eq.~\eqref{eq:g1d_analytical_expression} provides a reasonably accurate estimate. 

We remark here that Eq.~\eqref{eq:g1d_analytical_expression} decomposes the interatomic coupling constant into two contributions: the first ($\propto R_{\rm TF}$) corresponds to an ideal infinitely sharp barrier, while the second ($\propto s_x$) accounts for the finite width of the Gaussian barrier.


\subsection{Derivation of Equation (8)}

On the left rarefaction boundary we have zero local velocity \(u(x_l,t)=0\) and TF density $\rho(x_l,t)=s_0^2-\tfrac12\,\omega_x^2\,x_l^2,$
so that 
\begin{equation}
    r_\pm (x_l,t)=\pm \sqrt{s_0^2-\tfrac12\,\omega_x^2\,x_l^2},
\end{equation}
and hence \(r_+=-r_-\).

The characteristic speeds from Eq.~\eqref{eq:dispersionless_diagonal_system} are
\begin{equation}
  v_\pm = \frac{1}{2}(3\,r_\pm + r_\mp),  
\end{equation}
which, under \(r_-=-r_+\), reduce to
\(
v_+ = r_+,\,
v_- = -r_+.
\)
The left-moving solution is given by the slow branch \(v_-\).

Setting $v_- = \frac{\mathrm{d} x_l}{\mathrm{d}t}$ yields the initial value problem
\begin{equation}
    \frac{\mathrm{d} x_l}{\mathrm{d}t} = -\sqrt{s_0^2-\tfrac12\,\omega_x^2\,x_l^2},
\quad
x_l(0)=0.
\end{equation}
whose solution is Eq.~\eqref{eq:xl}.


\subsection{Matched solution}

Suppressing explicit time dependence for clarity, the coefficients appearing in Eq.~\eqref{eq:matched_solutions} are given by:
\begin{equation}\label{eq:matched_sol_coeffs}
\begin{aligned}
c_0 &= -(\alpha - c_2 x_r)x_r, & u_0 &= v_r - (\beta - u_2 x_r)x_r,  \\
c_1 &= \alpha - 2c_2 x_r, & u_1 &= \beta - 2u_2 x_r, \\
c_2 &= \frac{\alpha}{x_r - x_l} + \frac{C_l}{(x_r - x_l)^2}, & u_2 &= \frac{\beta}{x_r - x_l} - \frac{v_r}{(x_r - x_l)^2},
\end{aligned}
\end{equation}
with
\begin{equation}
  v_r(t) = 2 s_0 \cos(\omega_x t), \quad C_l(t) = \sqrt{\frac{g_{1\rm d}}{m}}\,|\psi_{\rm TF}(x_l(t))|.  
\end{equation}
Here, $|\psi_{\rm TF}(x)|$ is the TF density profile given by Eq.~\eqref{supp:TF_profile}, and $x_l(t)$ is the left edge defined in Eq.~\eqref{eq:xl}. The time-dependent coefficients $\alpha(t)$ and $\beta(t)$ are given by
\begin{equation}
  \alpha (t) = -\frac{\omega_x}{2} \csc\Big(\frac{3\omega_x t}{2}\Big), \quad
\beta (t) = \omega_x \cot\Big(\frac{3\omega_x t}{2}\Big).  
\end{equation}
The matched solution agrees well with the 1D and 3D computations up to roughly~\(100\)~ms. For longer times, the matched solution, being quadratic in space, does not accurately capture the dynamics, due to the formation of large gradients in the hydrodynamics quantities.

\end{document}